\documentclass[10pt,conference]{IEEEtran}

\usepackage{graphicx}
\usepackage{comment}
\usepackage{booktabs}
\usepackage{float}
\usepackage{adjustbox}
\usepackage{lscape}
 \usepackage{multirow}
 \usepackage[normalem]{ulem}
 \useunder{\uline}{\ul}{}
 \usepackage{lscape}
 \usepackage{longtable}
 \usepackage{array}
 \usepackage{url}
\usepackage{algorithm}
\usepackage[noend]{algpseudocode}
\usepackage{comment}
\usepackage{float}
\usepackage{adjustbox}
\usepackage{lscape}
 \usepackage{multirow}
 \usepackage[normalem]{ulem}
 \useunder{\uline}{\ul}{}
 \usepackage{lscape}
 \usepackage{longtable}
 \usepackage{array}
\usepackage{multibib}
 \usepackage[normalem]{ulem}
\usepackage[dvipsnames]{xcolor}

 \useunder{\uline}{\ul}{}

\title{Designing Adaptive Developer-Chatbot Interactions: Context Integration, Experimental Studies, and Levels of Automation}

\author{\IEEEauthorblockN{Glaucia Melo} 
\IEEEauthorblockA{\textit{David R. Cheriton School of Computer Science} \\
\textit{University of Waterloo}\\
Waterloo, Canada \\
gmelo@uwaterloo.ca}
}

\begin{document}

\maketitle

\begin{abstract}

The growing demand for software developers and the increasing development complexity have emphasized the need for support in software engineering projects. This is especially relevant in light of advancements in artificial intelligence, such as conversational systems. A significant contributor to the complexity of software development is the multitude of tools and methods used, creating various contexts in which software developers must operate. Moreover, there has been limited investigation into the interaction between context-based chatbots and software developers through experimental user studies. Assisting software developers in their work becomes essential. In particular, understanding the context surrounding software development and integrating this context into chatbots can lead to novel insight into what software developers expect concerning these human-chatbot interactions and their levels of automation. In my research, I study the design of context-based adaptive interactions between software developers and chatbots to foster solutions and knowledge to support software developers at work. 
\end{abstract}


\section{Introduction}

Software development is a challenging task requiring developers to gather and process information from various sources constantly. According to a study by Ponzanelli et al. \cite{ponzanelli2017supporting}, this information is often highly heterogeneous, making it difficult for developers to access and integrate all the information they need to complete their tasks effectively. To tackle this complexity, software developers utilize various tools and methods daily \cite{MEYER2017}. Some examples of these tools include integrated development environments (IDEs), version control systems (such as Git), and project management tools (such as JIRA). Given the rapidly growing demand for software development skills and the increasingly complex nature of the field \cite{labour2021}, software developers must receive adequate support and resources to help them succeed. 

Chatbots are computer programs that simulate human conversation using natural language processing (NLP) and machine learning (ML) techniques. These techniques allow chatbots to understand and respond to users' requests in a human-like way. There has been increased attention on software engineering chatbots \cite{Abdellatif2022}. Chatbot solutions have supported software developers in many ways: MSABot, PerformoBot \cite{Assavakamhaenghan2021}, TaskBot, MSRBot, MSABot \cite{Abdellatif2022} and others. However, there is limited published data on the design methods to support developers using context-based chatbots to perform their tasks \cite{selamat2021chatbot,nguyen2018understanding} and how to cope with customized levels of automation in their developer-chatbot interactions \cite{Engsley1997,parasuraman2000model,steinhauser2009design}. 

It is essential to examine the impact of the software development context on the software development process and consider the influence of developers' perspectives on designing context-based chatbot tools. Additionally, it is vital to consider the effect of varying levels of automation on the design of these tools.

My research bridges this gap by enhancing the design of context-based adaptive chatbots and fostering knowledge to improve these tools for better software development support. Specifically, I investigate the integration of software development context into the development process using chatbots and the interactions between developers and chatbots. I perform this investigation relying on three main research questions: \\
\textbf{RQ1: What types of context have been identified by researchers in software development projects?} Motivated by the complexity in understanding the contextual factors that affect software development supported by chatbots, I started by understanding what context is in software development projects. \\
\textbf{RQ2: How well can a chatbot support software developers when they are executing development tasks?} Motivated by the lack of research on design methods to support developers using context-based chatbots to perform their tasks and the lack of information from users (developers) to extract features for chatbots, I performed empirical user studies to extract information from developers to be applied in context-based chatbots. \\
\textbf{RQ3: Which factors impact the variability of levels of automation in autonomous systems?} Motivated by the lack of understanding of how to cope with customized levels of automation in human-chatbot interactions, I explore factors that influence levels of automation during human-systems interactions. Specifying the context in software development projects and studying the integration of context into the design of the adaptive interactions between software developers and chatbots can lead to a novel understanding of how to systemize context in software engineering and what software developers expect concerning context-based chatbot interactions and their level of automation.

\section{Background Information}

\subsection{Context in software engineering} As a complex knowledge-intensive effort \cite{Ciccio2015,MEYER2017}, many different contexts and technologies are involved in software development. In software engineering, context refers to the information and data surrounding and influencing a specific software system or application \cite{Murphy_Beyond2019}.  Context can include user input, system configuration, and other data the software needs to function correctly. It can also include the current state of the system, the current user, and any other information relevant to the software's operation. Context is essential in software engineering because it can influence the behavior and functionality of a software system. It can affect how the software processes input, generates output and interacts with other systems and users (developers). Understanding and managing context are crucial for building reliable, efficient, and easy software \cite{antunes2011context,murphy2018need, Murphy_Beyond2019,vilela2016systematic}. Furthermore, as a human-centred task \cite{Vasanthapriyan2015,Ponza2014Prompter}, software development fosters diverse practices, depending on each software developer's expertise, personal interests, gender \cite{lee2019floss,imtiaz2019investigating}, stress management \cite{Sarker2019}, and a variety of other user-related contexts of software development. Each software developer's technique when executing their tasks is heavily influenced by these many traits and the situation in which the developers work. A substantial corpus of research addresses the context of software development \cite{Kersten_Murphy_2006,Gasparic_Murphy_Ricci_2017,Holmes_Murphy_2005,MEYER2017}. The importance of treating these contexts as a first-class construct has been highlighted \cite{Murphy_Beyond2019} and can lead to significant changes in how developers execute their work.

Despite substantial research, the context in software development is still not explicit \cite{Murphy_Beyond2019}, nor is it presented as a framework that broadly supports the situation surrounding software developers. As a result, the ability to reuse this rich context across a software development project is severely limited. Furthermore, this context can be implicit and stored in the developers' minds (tacit) or dispersed through massive amounts of documentation \cite{Ciccio2015}. As they work on software projects, developers must maintain mental models of various tasks, and information \cite{LaToza06}. As a result, context can be easily lost or forgotten. With no history to infer contextualized information, developers are not supported by their current context \cite{avila2020}.
My suggestion for using chatbots in software development is to allow developers to communicate with a system designed to support them during development via conversations in which the chatbot should be aware of the context. The context-based chatbot should understand and capture contexts, such as the developer's repository, projects, and commands. This motivates the continuation of my research as described next.

\subsection{Chatbots to support software development} According to Murphy et al. \cite{Murphy_Beyond2019}, utilizing context with intelligent assistants, like chatbots, could allow developers to focus on the challenging aspects of problems while the assistant does routine activities. Bots are used in software development to serve a variety of activities, from automating tedious chores to addressing knowledge and communication gaps in software teams \cite{storey2016disrupting}. Other works have focused on different aspects of the use of chatbots in the realm of software engineering. Researchers focused on detecting code conflicts when developers work in parallel \cite{paikari2019chatbot} or recommending the right person to contact in open source projects \cite{cerezo2019building}.  According to recent research, 26\% of OSS projects on GitHub use software bots for various activities \cite{mairieli2018}. However, there's still a lack of research that addresses developers' inputs on context-based chatbots with real scenarios \cite{Assavakamhaenghan2021}. Little is known about software developers' perceptions of using chatbots at work \cite{Assavakamhaenghan2021, Abdellatif2022}. This indicates a need to understand the perceptions of context-aware chatbots among software developers and to empirically explore design opportunities for such tools, triangulating current and future solutions with design requirements. 

\subsection{Levels of Automation} Machines have taken over many functions and operations that were formerly entirely conducted by humans \cite{ford2015rise,mikic2021impact}. Examples are manufacturing and assembly line jobs (welding, painting, packaging), data entry, bookkeeping, other administrative tasks, customer service, such as answering phones and responding to emails, and many others  are now computerized. The advantages of this increased automation are numerous, including enhanced productivity, quality improvement, accuracy, and others \cite{VAGIA2016190}. Artificial intelligence (AI) advancements have recently significantly impacted how organizations and society operate, with machines taking over human responsibilities. Many debate the benefits and drawbacks of full automation against keeping humans involved in specific tasks \cite{flemisch2012towards}. Autonomous systems can be built and deployed to best match human and machine capabilities. One approach is to designate distinct levels of automation (LOA) for different jobs \cite{VAGIA2016190, endsley1999level}. Although research in manufacturing and robot interaction exists \cite{wang2020overview}, there is a lack of research that aims at understanding the factors that influence human-system collaboration so that adaptive allocation of tasks in different levels of automation can be provided \cite{Engsley1997,parasuraman2000model,steinhauser2009design}. 

\section{Methodology and Progress}

Ideally, contextual information (domain, environment, tools, human aspects) should be understood so developers can receive contextual support. Consequently, this motivates a need to investigate this context (RQ1) and mechanisms (tools, approaches) in which this contextual support can be delivered to developers (RQ2, RQ3). My dissertation relies on a mixed-methods approach to address the proposed research questions. Next, I present the details of the methodologies used and the progress of my research.  

\subsection{RQ1: What types of context have been identified by researchers in software development projects?}

To address RQ1, I conducted a systematic literature review (SLR) \cite{kitchenham2007guidelines} on contexts in software development, supported by nine sub-questions. As a result, I (1) highlighted the findings of the SLR and consolidated the results found in the literature in a table, and (2) suggested, based on the SLR results, a preliminary context-augmented framework for software development projects. This proposed framework allows for capturing and monitoring the context variabilities. Aside from these two contributions, I also (3) proposed a context model with the contexts found in the SLR. With the representation benefits of a model, this context model exposes the variability of contexts observed in the literature review (environment, people, domain) and how these contexts could interact. This research was published in 2019 \cite{melo2019context}. Understanding and integrating context in software development allows for significant improvement in how tools can aid software developers \cite{Murphy_Beyond2019}.

Essentially, this work has shown extensive contexts surrounding software development. While many academics studied the collection and reuse of software development context \cite{Kersten_Murphy_2006,Gasparic_Murphy_Ricci_2017,Holmes_Murphy_2005,MEYER2017, LaToza06, Sawadsky2011}, we still need an approach that integrates this variable context into the software development so that developers can receive tailored information according to contextual changes in real-time. One study claims software developers use around 15 applications per day \cite{MEYER2017}. Existing tools or methodologies must be prepared to respond well to contextual changes \cite{nascimento2018iot}. 

One solution to communicate (gather and present) context to software developers and support their work is through chatbots \cite{devy}. Therefore, I propose discussing how to make this solution using context-based chatbots viable. This discussion was published at the International Conference in Software Engineering (ICSE 2021) as a New Ideas and Emerging Results (NIER) paper \cite{melo2021cognitive}. This paper outlines the required future work to build the robust proposed approach. As this strategy mainly relies on chatbots as a support tool during software development (SD), as inspired by the work of Bradley et al. \cite{devy}, I chose to study how interested software engineers were in using chatbots at work, as well as empirically uncover  design opportunities are there for such tools. Hence, I conducted the next set of studies. 

\subsection{RQ2: How well can chatbots support developers?}
To answer this RQ, I conducted a user study investigating developers' perceptions and design opportunities for context-based chatbots for software development. I started by running a pilot study using the Wizard-of-Oz methodology in a classroom scenario \cite{melo2020exploring}. This study analyzed whether supporting software developers with a chatbot during task execution can improve the overall development experience. I ran this study with five graduate student colleagues with some software development experience. 

Because the results were promising but limited in terms of methodology and the number and experience of participants, I ran a within-subjects study with 29 participants, using a chatbot prototype and applying a robust methodology. I thereby brought the lessons learned from the pilot study to a new user study involving software practitioners as participants. In this new study, the chatbot aims to support developers with software development information, recommending or executing task workflows. I empirically captured developers' preferences for task support and opinions on whether the chatbot can support developers during the execution of current project tasks rather than upholding developers' memory (e.g., through mind maps). 

In this study, I presented participants (29 software developers) with a scenario which described a day in the life of a software developer. Then, participants would interact with Devbot (the prototype), where they were supposed to ask questions that arose when reading the scenario. Devbot was only ready to answer the questions about the scenario, which were targeted at specific attributes of contexts, such as tasks available, IDE tool interaction, repository management, agile methods and team collaboration;
any other question would result in the utterance "Sorry, I can not help you with that. Is there anything else I can help with?", to prompt developers to keep asking questions. Then, the participants would fill out a questionnaire and respond to interview questions. 

In effect, I investigate five main topics: (1) if having a context-based chatbot to support their daily work, if developers are willing to use it, (2) what types of tasks this chatbot should support, (3) what types of questions should chatbots be able to answer, (4) if some unexpected questions/requests were not anticipated, and finally, (5) the overall opinion of the developers regarding the use of a context-based chatbot and how questions are managed, and what could increase the tool adoption. Preliminary findings show (1) developers are interested in adopting chatbots as a support tool during software development, with (2) particular emphasis on using the chatbot for issue management, in comparison to all the options presented in the scenario (code repository management, editing a java class, looking for further instructions on how to solve the task at hand). Findings also show that (3) developers are eager to work with such tools and find this solution intriguing while also (4) presenting design opportunities for how chatbots might act and communicate and which features they would like to see (seeing common causes of merge conflicts and resources, setting priorities for tasks, and others). One last intriguing finding was that while some participants expressed enjoying high levels of automation (the chatbot executing tasks autonomously), others preferred to be guided through tasks they would execute themselves. Preliminary results of this study were published in 2020 \cite{melo_20_understanding}. The study used a prototype and interviews with 29 developers from diverse backgrounds; limitations acknowledged and addressed by selecting participants from different demographic groups. Data limitations and biases will be addressed with multiple coders in the new data analysis. Researcher bias was minimized through scenario checking and referencing other papers.

Given the substantial data I collected in this study (a rich list of over 500 questions, the questionnaire results and the interview transcriptions), I intend to extract further information from this data. Therefore, keeping the original goal of empirically informing design opportunities for chatbots and understanding the interaction between developers and chatbots, from the developers' perspective, in 2023, I will perform a second data analysis over the same study data. In this second analysis round, I will look for the format patterns in the questions asked and more specific usability suggestions from software developers from the interviews and questionnaire text. Therefore, this second analysis will not be related to the chatbot's capabilities in terms of the topics of interest to software developers but rather to assess this data and guide software engineering-related context-based chatbot design, triangulating current studies and tools developed to support software developers with empirical results and providing a set of actionable recommendations for software development chatbot creators. I will perform the following analyses:

\begin{itemize}
\item Quantitative: How many questions were asked, for how many minutes in total participants interacted with the chatbot and how many questions were related to the scenario and how many were not related
\item Qualitative: Perform Open Coding or Thematic Network Analysis (TNA) to extract categories from interview transcriptions on the topics of the emerging concerns of software developers when interacting with context-based chatbots and emerging design opportunities. Use a Correlation Matrix to investigate data correlations such as age vs level of automation, experience vs levels of automation and other possible correlations from the data collected. I plan to make all data anonymized and available for research purposes. I intend to finalize with a triangulation of the design opportunities extracted in this study, by comparing them with existing features in current chatbot tools such as GitHub's Hubot, Botkit, StackStorm and others
\end{itemize}

\subsection{RQ3: Which factors impact the variability of levels of automation in autonomous systems?}

To gain a deeper understanding of the factors that impact the level of automation in human-system interactions and how these levels of automation affect the design of such interactions, I have conducted the next study. This study includes a systematic literature review to identify and summarize the key factors that determine the level of automation in human-system interactions. I then represent these factors and their relationships in a feature model and with notations. To support the development of systems that adapt their functions according to levels of automation, modelling and representing the factors that influence how systems will adapt is essential. This study adds to the body of knowledge by providing a feature-model-based depiction of the factors affecting the degree of automation in autonomous systems. The present results are significant in at least two major respects. First, it highlights the existence of these factors, categorizes them and presents how a combination of different factors can influence how intelligent autonomous systems work. We also demonstrate how systems can capture factors and systemize their use. Finally, we achieve system adaptability and characterization by identifying and representing these factors as feature models and evaluating the models with case studies. 

In this study, I expose five main categories of factors that influence levels of automation: Quality, Human, Environment, System and Task factors. The current findings are significant because they emphasize the existence of these factors and provide designers with references to which aspects potentially influence levels of automation in human-system interactions. As a result, systems can adapt how much automation is used based on factors determining this degree. The preliminary results of this study were published in 2022, and another paper is currently under submission to a major academic journal. 

\section{Contributions}
In my research project, I tackle the problem of integrating contextual software development information into development through context-based chatbots. Specifically by understanding what this context is, what software developers expect concerning context-based chatbot interactions, and the level of automation desired. This way, we increase the body of knowledge on context-based chatbot approaches to support software development and foster the understanding of the design of context-based adaptive chatbots so that these tools, once built, can provide optimal support to software developers.
A summary of contributions of my Ph.D. research are: 
\begin{itemize}
\item Within-subjects user study and experimental evaluation
\begin{itemize}
\item Publicly available user study data
\end{itemize}
\item Approach and results of a systematic literature review revealing contextual factors in software engineering 
\item Context model design proposal
\item Context-based chatbot design
\item Approach and results of a systematic literature review revealing factors that influence levels of automation (LOAs), categorization of these factors, and the results of the analysis of the relationships between these identified factors and LOAs
\item Design model that demonstrates the variability of the factors and LOA using feature models and constraints 
\item Use cases in two different domains and scenarios to evaluate feature models
\end{itemize}

\section{Timeline for Completion}
I will conclude the planned further data analysis of the user study by March 2023. I plan to get feedback and organize my results into a dissertation by June 2023 and  defend my doctoral thesis by September 2023.

\bibliographystyle{IEEEtran} 
\bibliography{main.bib,loa.bib}

\begin{thebibliography}{10}
\providecommand{\url}[1]{#1}
\csname url@samestyle\endcsname
\providecommand{\newblock}{\relax}
\providecommand{\bibinfo}[2]{#2}
\providecommand{\BIBentrySTDinterwordspacing}{\spaceskip=0pt\relax}
\providecommand{\BIBentryALTinterwordstretchfactor}{4}
\providecommand{\BIBentryALTinterwordspacing}{\spaceskip=\fontdimen2\font plus
\BIBentryALTinterwordstretchfactor\fontdimen3\font minus
  \fontdimen4\font\relax}
\providecommand{\BIBforeignlanguage}[2]{{%
\expandafter\ifx\csname l@#1\endcsname\relax
\typeout{** WARNING: IEEEtran.bst: No hyphenation pattern has been}%
\typeout{** loaded for the language `#1'. Using the pattern for}%
\typeout{** the default language instead.}%
\else
\language=\csname l@#1\endcsname
\fi
#2}}
\providecommand{\BIBdecl}{\relax}
\BIBdecl

\bibitem{ponzanelli2017supporting}
L.~Ponzanelli, S.~Scalabrino, G.~Bavota, A.~Mocci, R.~Oliveto, M.~Di~Penta, and
  M.~Lanza, ``Supporting software developers with a holistic recommender
  system,'' in \emph{2017 IEEE/ACM 39th International Conference on Software
  Engineering (ICSE)}.\hskip 1em plus 0.5em minus 0.4em\relax IEEE, 2017, pp.
  94--105.

\bibitem{MEYER2017}
A.~N. Meyer, L.~E. Barton, G.~C. Murphy, T.~Zimmermann, and T.~Fritz,
  ``\BIBforeignlanguage{English}{The work life of developers: Activities,
  switches and perceived productivity},''
  \emph{\BIBforeignlanguage{English}{IEEE Transactions on Software
  Engineering}}, vol.~43, no.~12, pp. 1178--1193, 2017.

\bibitem{labour2021}
\BIBentryALTinterwordspacing
``Bureau of labor statistics, u.s. department of labor, occupational outlook
  handbook: Software developers, quality assurance analysts, and testers,''
  accessed: 2022-11-06. [Online]. Available:
  \url{https://www.bls.gov/ooh/computer-and-information-technology/software-developers.htm}
\BIBentrySTDinterwordspacing

\bibitem{Abdellatif2022}
A.~Abdellatif, K.~Badran, D.~E. Costa, and E.~Shihab, ``A comparison of natural
  language understanding platforms for chatbots in software engineering,''
  \emph{IEEE Transactions on Software Engineering}, vol.~48, no.~8, pp.
  3087--3102, 2022.

\bibitem{Assavakamhaenghan2021}
N.~Assavakamhaenghan, R.~G. Kula, and K.~Matsumoto, ``Interactive chatbots for
  software engineering: A case study of code reviewer recommendation,'' in
  \emph{2021 IEEE/ACIS 22nd International Conference on Software Engineering,
  Artificial Intelligence, Networking and Parallel/Distributed Computing
  (SNPD)}, 2021, pp. 262--266.

\bibitem{selamat2021chatbot}
M.~A. Selamat and N.~A. Windasari, ``Chatbot for smes: Integrating customer and
  business owner perspectives,'' \emph{Technology in Society}, vol.~66, p.
  101685, 2021.

\bibitem{nguyen2018understanding}
Q.~N. Nguyen and A.~Sidorova, ``Understanding user interactions with a chatbot:
  A self-determination theory approach,'' \emph{24th Americas Conference on
  Information Systems – Emergent Research Forum}, 2018.

\bibitem{Engsley1997}
M.~Engsley, E.~Onal, and D.~Kaber, ``The impact of intermediate levels of
  automation on situation awareness and performance in dynamic control
  systems,'' in \emph{Proceedings of the 1997 IEEE Sixth Conference on Human
  Factors and Power Plants, 1997. 'Global Perspectives of Human Factors in
  Power Generation'}, 1997, pp. 7/7--712.

\bibitem{parasuraman2000model}
R.~Parasuraman, T.~B. Sheridan, and C.~D. Wickens, ``A model for types and
  levels of human interaction with automation,'' \emph{IEEE Transactions on
  systems, man, and cybernetics-Part A: Systems and Humans}, vol.~30, no.~3,
  pp. 286--297, 2000.

\bibitem{steinhauser2009design}
N.~B. Steinhauser, D.~Pavlas, and P.~A. Hancock, ``Design principles for
  adaptive automation and aiding,'' \emph{Ergonomics in Design}, vol.~17,
  no.~2, pp. 6--10, 2009.

\bibitem{Ciccio2015}
C.~Di~Ciccio, A.~Marrella, and A.~Russo,
  ``\BIBforeignlanguage{English}{Knowledge-intensive processes:
  Characteristics, requirements and analysis of contemporary approaches},''
  \emph{\BIBforeignlanguage{English}{Journal on Data Semantics}}, vol.~4,
  no.~1, pp. 29--57, 2015.

\bibitem{Murphy_Beyond2019}
G.~Murphy, ``Beyond integrated development environments: adding context to
  software development,'' in \emph{Proceedings of the 41st International
  Conference on Software Engineering}.\hskip 1em plus 0.5em minus 0.4em\relax
  IEEE Press, 2019, pp. 73--76.

\bibitem{antunes2011context}
B.~Antunes, F.~Correia, and P.~Gomes, ``Context capture in software
  development,'' \emph{arXiv preprint arXiv:1101.4101}, 2011.

\bibitem{murphy2018need}
G.~C. Murphy, ``The need for context in software engineering (ieee cs harlan
  mills award keynote),'' in \emph{33rd IEEE/ACM International Conference on
  Automated Software Engineering (ASE)}, 2018, pp. 5--5.

\bibitem{vilela2016systematic}
J.~Vilela, J.~Castro, and J.~Pimentel, ``A systematic process for obtaining the
  behavior of context-sensitive systems,'' \emph{Journal of Software
  Engineering Research and Development}, vol.~4, pp. 1--57, 2016.

\bibitem{Vasanthapriyan2015}
S.~Vasanthapriyan, J.~Tian, and J.~Xiang, ``A survey on knowledge management in
  software engineering,'' in \emph{Software Quality, Reliability and
  Security-Companion (QRS-C), 2015 IEEE International Conference on}.\hskip 1em
  plus 0.5em minus 0.4em\relax IEEE, 2015, pp. 237--244.

\bibitem{Ponza2014Prompter}
L.~Ponzanelli, G.~Bavota, M.~Di~Penta, R.~Oliveto, and M.~Lanza, ``Mining
  stackoverflow to turn the ide into a self-confident programming prompter,''
  in \emph{Proceedings of the 11th working conference on mining software
  repositories}, 2014, pp. 102--111.

\bibitem{lee2019floss}
A.~Lee and J.~C. Carver, ``Floss participants' perceptions about gender and
  inclusiveness: a survey,'' in \emph{Proceedings of the 41st International
  Conference on Software Engineering}.\hskip 1em plus 0.5em minus 0.4em\relax
  IEEE Press, 2019, pp. 677--687.

\bibitem{imtiaz2019investigating}
N.~Imtiaz, J.~Middleton, J.~Chakraborty, N.~Robson, G.~Bai, and E.~Murphy-Hill,
  ``Investigating the effects of gender bias on github,'' in \emph{Proceedings
  of the 41st International Conference on Software Engineering}.\hskip 1em plus
  0.5em minus 0.4em\relax IEEE Press, 2019, pp. 700--711.

\bibitem{Sarker2019}
F.~Sarker, B.~Vasilescu, K.~Blincoe, and V.~Filkov, ``Socio-technical work-rate
  increase associates with changes in work patterns in online projects,'' in
  \emph{Proceedings of the 41st International Conference on Software
  Engineering}.\hskip 1em plus 0.5em minus 0.4em\relax IEEE Press, 2019, pp.
  936--947.

\bibitem{Kersten_Murphy_2006}
M.~Kersten and G.~C. Murphy, ``Using task context to improve programmer
  productivity,'' in \emph{Proceedings of the 14th ACM SIGSOFT International
  Symposium on Foundations of Software Engineering}, ser. SIGSOFT
  ’06/FSE-14.\hskip 1em plus 0.5em minus 0.4em\relax ACM, 2006, p. 1–11.

\bibitem{Gasparic_Murphy_Ricci_2017}
M.~Gasparic, G.~C. Murphy, and F.~Ricci, ``A context model for ide-based
  recommendation systems,'' \emph{Journal of Systems and Software}, vol. 128,
  p. 200–219, Jun 2017.

\bibitem{Holmes_Murphy_2005}
R.~Holmes and G.~C. Murphy, ``Using structural context to recommend source code
  examples,'' in \emph{Proceedings of the 27th International Conference on
  Software Engineering}, ser. ICSE ’05.\hskip 1em plus 0.5em minus
  0.4em\relax ACM, 2005, p. 117–125, event-place: St. Louis, MO, USA.

\bibitem{LaToza06}
\BIBentryALTinterwordspacing
T.~D. LaToza, G.~Venolia, and R.~DeLine, ``Maintaining mental models: A study
  of developer work habits,'' in \emph{Proceedings of the 28th International
  Conference on Software Engineering}, ser. ICSE '06.\hskip 1em plus 0.5em
  minus 0.4em\relax New York, NY, USA: Association for Computing Machinery,
  2006, p. 492–501. [Online]. Available:
  \url{https://doi.org/10.1145/1134285.1134355}
\BIBentrySTDinterwordspacing

\bibitem{avila2020}
L.~F. D'Avila, J.~L.~V. Barbosa, and K.~S.~F. de~Oliveira, ``Sw-context: a
  model to improve developers’ situational awareness,'' \emph{IET Software},
  vol.~14, no.~5, pp. 535--543, 2020.

\bibitem{storey2016disrupting}
M.-A. Storey and A.~Zagalsky, ``Disrupting developer productivity one bot at a
  time,'' in \emph{Proceedings of the 2016 24th ACM SIGSOFT International
  Symposium on Foundations of Software Engineering}, 2016, pp. 928--931.

\bibitem{paikari2019chatbot}
E.~Paikari, J.~Choi, S.~Kim, S.~Baek, M.~Kim, S.~Lee, C.~Han, Y.~Kim, K.~Ahn,
  C.~Cheong \emph{et~al.}, ``A chatbot for conflict detection and resolution,''
  in \emph{2019 IEEE/ACM 1st International Workshop on Bots in Software
  Engineering (BotSE)}.\hskip 1em plus 0.5em minus 0.4em\relax IEEE, 2019, pp.
  29--33.

\bibitem{cerezo2019building}
J.~Cerezo, J.~Kubelka, R.~Robbes, and A.~Bergel, ``Building an expert
  recommender chatbot,'' in \emph{2019 IEEE/ACM 1st International Workshop on
  Bots in Software Engineering (BotSE)}.\hskip 1em plus 0.5em minus 0.4em\relax
  IEEE, 2019, pp. 59--63.

\bibitem{mairieli2018}
\BIBentryALTinterwordspacing
M.~Wessel, B.~M. de~Souza, I.~Steinmacher, I.~S. Wiese, I.~Polato, A.~P.
  Chaves, and M.~A. Gerosa, ``The power of bots: Characterizing and
  understanding bots in oss projects,'' \emph{Proceedings of the ACM on
  Human-Computer Interaction}, vol.~2, no. CSCW, Nov. 2018. [Online].
  Available: \url{https://doi.org/10.1145/3274451}
\BIBentrySTDinterwordspacing

\bibitem{ford2015rise}
M.~Ford, \emph{Rise of the Robots: Technology and the Threat of a Jobless
  Future}.\hskip 1em plus 0.5em minus 0.4em\relax Basic Books, 2015.

\bibitem{mikic2021impact}
M.~Mikic and J.~Malala, ``The impact of artificial intelligence on the future
  of work,'' in \emph{The Home in the Digital Age}, 2021, pp. 143--159.

\bibitem{VAGIA2016190}
M.~Vagia, A.~A. Transeth, and S.~A. Fjerdingen, ``A literature review on the
  levels of automation during the years. what are the different taxonomies that
  have been proposed?'' \emph{Applied Ergonomics}, vol.~53, pp. 190--202, 2016.

\bibitem{flemisch2012towards}
F.~Flemisch, M.~Heesen, T.~Hesse, J.~Kelsch, A.~Schieben, and J.~Beller,
  ``Towards a dynamic balance between humans and automation: authority,
  ability, responsibility and control in shared and cooperative control
  situations,'' \emph{Cognition, Technology \& Work}, vol.~14, no.~1, pp.
  3--18, 2012.

\bibitem{endsley1999level}
M.~R. Endsley and D.~B. Kaber, ``Level of automation effects on performance,
  situation awareness and workload in a dynamic control task,''
  \emph{Ergonomics}, vol.~42, no.~3, pp. 462--492, 1999.

\bibitem{wang2020overview}
L.~Wang, S.~Liu, H.~Liu, and X.~V. Wang, ``Overview of human-robot
  collaboration in manufacturing,'' in \emph{Proceedings of 5th International
  Conference on the Industry 4.0 Model for Advanced Manufacturing: AMP
  2020}.\hskip 1em plus 0.5em minus 0.4em\relax Springer, 2020, pp. 15--58.

\bibitem{kitchenham2007guidelines}
B.~Kitchenham and S.~Charters, ``Guidelines for performing systematic
  literature reviews in software engineering,'' \emph{Technical Report EBSE
  2007-001}, 2007.

\bibitem{melo2019context}
G.~Melo, P.~Alencar, and D.~Cowan, ``Context-augmented software development in
  traditional and big data projects: Literature review and preliminary
  framework,'' in \emph{2019 IEEE International Conference on Big Data (Big
  Data)}.\hskip 1em plus 0.5em minus 0.4em\relax IEEE, 2019, pp. 3449--3457.

\bibitem{Sawadsky2011}
N.~Sawadsky and G.~C. Murphy, ``Fishtail: From task context to source code
  examples,'' in \emph{Proceedings of the 1st Workshop on Developing Tools as
  Plug-Ins}, ser. TOPI '11.\hskip 1em plus 0.5em minus 0.4em\relax New York,
  NY, USA: Association for Computing Machinery, 2011, p. 48–51.

\bibitem{nascimento2018iot}
N.~Nascimento, P.~Alencar, C.~Lucena, and D.~Cowan, ``An iot analytics embodied
  agent model based on context-aware machine learning,'' in \emph{2018 IEEE
  International Conference on Big Data (Big Data)}.\hskip 1em plus 0.5em minus
  0.4em\relax IEEE, 2018, pp. 5170--5175.

\bibitem{devy}
N.~Bradley, T.~Fritz, and R.~Holmes, ``Context-aware conversational developer
  assistants,'' in \emph{2018 IEEE/ACM 40th International Conference on
  Software Engineering (ICSE)}.\hskip 1em plus 0.5em minus 0.4em\relax IEEE,
  2018, pp. 993--1003.

\bibitem{melo2021cognitive}
G.~Melo, P.~Alencar, and D.~Cowan, ``A cognitive and machine learning-based
  software development paradigm supported by context,'' in \emph{2021 IEEE/ACM
  43rd International Conference on Software Engineering: New Ideas and Emerging
  Results (ICSE-NIER)}, 2021, pp. 11--15.

\bibitem{melo2020exploring}
G.~Melo, E.~Law, P.~Alencar, and D.~Cowan, ``Exploring context-aware
  conversational agents in software development,'' \emph{CoRR}, vol.
  abs/2006.02370, 2020.

\bibitem{melo_20_understanding}
------, ``Understanding user understanding: What do developers expect from a
  cognitive assistant?'' in \emph{2020 IEEE International Conference on Big
  Data (Big Data)}.\hskip 1em plus 0.5em minus 0.4em\relax IEEE Computer
  Society, 2020, pp. 3165--3172.

\end{thebibliography}

\end{document}